%% file: main.tex
\documentclass[journal,twoside,web]{ieeecolor2}

\makeatletter
\let\ps@titlepagestyle\ps@plain
\makeatother

\usepackage{generic}
\usepackage{cite}
\usepackage{amsmath,amssymb,amsfonts}
\usepackage{graphicx}
\usepackage{float}
\usepackage{textcomp}
\usepackage{booktabs}
\usepackage{siunitx}
\usepackage{stfloats}
\usepackage{subcaption}
\usepackage[font=small]{caption}
\usepackage[hidelinks,hypertexnames=false]{hyperref}

\usepackage{tabularx}
\usepackage{makecell}
\newcolumntype{Y}{>{\raggedright\arraybackslash}X}

\title{VideoPulse: Neonatal Heart Rate and SpO\textsubscript{2} Estimation from Contact Free Video}

\author{Deependra Dewagiri,~Kamesh Anuradha,~Pabadhi Liyanage,~Helitha Kulatunga,~Pamuditha Somarathne,~\IEEEmembership{Graduate Student Member,~IEEE},~Udaya S. K. P. Miriya~Thanthrige,~\IEEEmembership{Member,~IEEE},~Nishani Lucas,~Anusha Withana,~\IEEEmembership{Member,~IEEE}, and Joshua P. Kulasingham,~\IEEEmembership{Member,~IEEE}
\thanks{This research was
supported by the Australian Government through the Australian
Research Council Future Fellowship under Grant FT250100813,
awarded to A. Withana. (Corresponding authors: Joshua Pranjeevan Kulasingham and Deependra Dewagiri.)}%
\thanks{This work involved human subjects in its research.
Approval of all ethical and experimental procedures and protocols
was granted by the University Ethics Review Committee, University
of Moratuwa, under Ethics Declaration No. ERN/2024/004, and by the
Institutional Ethical Review Committee of De Soysa Maternity
Hospital for Women under Application No. 026/2024. Written informed
consent was obtained from a parent or legal guardian of each neonatal
participant before enrollment and data collection.}%
\thanks{D. Dewagiri, K. Anuradha, P. Liyanage, H. Kulatunga,
U. S. K. P. M. Thanthrige, and J. P. Kulasingham are with the
Department of Electronic and Telecommunication Engineering,
University of Moratuwa, Moratuwa 10400, Sri Lanka
(e-mail: pranjeevank@uom.lk; deependradewagiri@gmail.com).}%
\thanks{P. Somarathne and A. Withana are with the School of
Computer Science, Faculty of Engineering, The University of Sydney,
Sydney, NSW 2006, Australia.}%
\thanks{N. Lucas is with the Department of Paediatrics, Faculty
of Medicine, University of Colombo, Colombo 08, Sri Lanka.}%
}

\begin{document}
\maketitle

\thispagestyle{plain}
\pagestyle{plain}

\input{abstract}


\input{introduction}

\input{related_work}
\input{methodology}

\input{datasets}

\input{results}

\input{discussion}

\input{conclusion}
\input{acknowledgement}

\bibliographystyle{IEEEtran}
\bibliography{references}

\end{document}

%% file: abstract.tex
\begin{abstract}
\textit{Objective:} Remote photoplethysmography (rPPG) allows for contactless vital sign monitoring, which is beneficial for sensitive neonates. We aim to develop an rPPG pipeline to estimate neonatal heart rate (HR) and peripheral capillary oxygen saturation (SpO\textsubscript{2}) from facial video. \textit{Methods:} We propose a novel deep learning based rPPG pipeline, VideoPulse, that extends a three-dimensional convolutional neural network for HR estimation to SpO\textsubscript{2} regression from facial video. Our pipeline combines face alignment, a generative adversarial network for reference signal denoising, and class imbalance handling. Video recordings from 52 neonates were collected at varied orientations. Our pipeline was evaluated on the prior NBHR neonatal dataset and on our own dataset. \textit{Results:} On NBHR, using short 2 s windows, mean absolute errors were 2.97 bpm for HR and 1.69\% for SpO\textsubscript{2}. On our own dataset, the NBHR-trained HR model achieved 5.34 bpm without fine-tuning; the NBHR-pretrained SpO\textsubscript{2} model achieved 1.68\% after fine-tuning on our dataset. \textit{Conclusion:} Our work proves that neonatal HR and SpO\textsubscript{2} estimation can be reliably achieved from short facial video segments. \textit{Significance:} This work proposes a novel rPPG pipeline, validated across two datasets, and lays the foundation for contactless vital sign monitoring of neonates under realistic situations.
\end{abstract}

%% file: introduction.tex
\section{Introduction}

Physiological signals such as heart rate (HR), respiratory rate (RR), and peripheral capillary oxygen saturation (SpO\textsubscript{2}) are important indicators in clinical assessment and continuous patient monitoring. Electrocardiography (ECG) and contact photoplethysmography (PPG), including pulse oximetry, are widely used to measure these signals. Remote photoplethysmography (rPPG) extends PPG to standard cameras by tracking subtle intensity variations in facial video \cite{autohr, TSCAn}. These variations arise from blood volume dynamics and respiration, enabling contact free estimation of vital signs such as HR and SpO\textsubscript{2} \cite{cheng2024contactless}, which is useful when contact sensors are impractical or increase infection risk \cite{TSCAn}. In neonatal care, adhesive electrodes and prolonged probe contact can irritate fragile skin \cite{app11167215}, motivating contact free alternatives for monitoring in the Neonatal Intensive Care Unit (NICU) and for follow up after discharge.

Early research on rPPG primarily relied on signal processing methods that analyze subtle color changes in facial regions of interest (ROI) to extract pulse related signals \cite{7565547}. In practice, rPPG signals are inherently weak and are easily corrupted by motion, pose changes, illumination variation, video compression, and sensor noise \cite{7565547}. These challenges limit the effectiveness of signal processing techniques in realistic clinical environments. To overcome these limitations, machine learning methods have been utilized which often result in improved accuracy and robustness \cite{SynRhythm, 9607515, electronics13071334, physnet, deepphys,10205146}.

Recent work explores architectures that model long temporal dependencies, including attention based models \cite{physformer}. However, their computational cost and sensitivity to sequence length can be challenging for real time bedside use. More efficient sequence modeling has recently been explored using selective state space models, such as Mamba \cite{mamba} and rPPG oriented adaptations like RhythmMamba \cite{RhythmMambaFR}. However, for neonatal monitoring, reliable inference from short windows (e.g., 2 s) and modest compute is particularly important, but current models fall short in these areas \cite{physformer}.


Most rPPG models have been developed and validated using adult datasets, with limited research conducted on neonates. HR inference from neonatal facial video is challenging because ward recordings often contain motion, occlusion, and illumination variation. 
Since rPPG performance can vary with capture conditions and subject appearance, it is important to assess models on diverse cohorts and realistic clinical environments \cite{Dasari2021}. However, the few neonatal rPPG studies report limited demographic and skin tone diversity \cite{neonates}, which restricts evaluation across broader patient populations. Furthermore, these few studies have focused primarily on HR estimation \cite{neonates}, and, to the best of our knowledge, neonatal SpO\textsubscript{2} estimation from facial video remains underexplored. 

To address these limitations, we propose an algorithmic pipeline \emph{VideoPulse} capable of estimating neonatal HR and SpO\textsubscript{2} from facial video and validate it on a prior neonatal dataset (NBHR \cite{neonates}). Our proposed pipeline uses state-of-the-art video based HR estimation methods and extends them for SpO\textsubscript{2} prediction, while incorporating techniques for alignment, signal denoising and addressing class imbalances. Additionally, we also collect our own neonatal dataset, with synchronized facial video, HR, SpO\textsubscript{2}, and pulse oximeter PPG reference signals. Compared with NBHR\cite{neonates}, which is the only widely used publicly available neonatal rPPG dataset, our dataset represents a population with a different skin tone and reflects realistic ward acquisition, including variation in lighting conditions and pose variation across three orientations and variable capture conditions (such as the baby lying on the cot or on the bed). 

Our contributions are threefold. First, we present an end to end system that estimates neonatal HR and SpO\textsubscript{2} from facial video under challenging clinical conditions, including illumination variation and unaligned faces. Second, we introduce a supervision and training strategy for SpO\textsubscript{2} estimation that combines pulse oximeter PPG denoising with label distribution smoothing and weighted regression to mitigate label imbalance. Third, we proved that our method surpassed state-of-the-art results by evaluating our pipeline on both a prior dataset and our own neonatal dataset which contains synchronized video, HR, SpO\textsubscript{2}, and pulse oximeter PPG reference. This work demonstrates the potential of contact-free vital sign monitoring to improve neonatal care by enabling accurate, non-invasive, and continuous assessment in clinical settings.

%% file: related_work.tex
\section{Related Work}

\subsection{Signal processing methods for rPPG}

Early rPPG research focused on recovering physiological signals from subtle color changes in facial skin regions using hand-crafted signal processing techniques. Widely used approaches include Plane Orthogonal to Skin (POS) and CHROM, which project RGB signals into chrominance spaces designed to suppress illumination variation while enhancing pulsatile components \cite{7565547,dehaan2013chrom,vanes2023contactless,haugg2022effectiveness,Dasari2021,acharya2025reliability}. Other methods apply blind source separation or decomposition techniques, such as Principal Component Analysis (PCA) and Independent Component Analysis (ICA), to isolate cardiac-related signals from mixed RGB observations \cite{lewandowska2011measuring,yu2015dynamic,wedekind2017assessment,poh2010noncontact,tsouri2012constrained,macwan2018remote}.

Although these methods can achieve reliable HR estimation under controlled settings, their performance often deteriorates in realistic environments. In neonatal care settings, challenges such as subject motion, pose variation, inconsistent illumination, partial facial visibility, and video compression artifacts can significantly distort weak pulsatile signals \cite{vipl}. Furthermore, most classical pipelines rely on carefully selected regions of interest and handcrafted assumptions about skin reflection characteristics, which may not generalize across diverse neonatal populations or clinical acquisition conditions.

\subsection{Learning-based rPPG methods}

To improve robustness under challenging conditions, recent work has increasingly adopted deep learning approaches for rPPG estimation. Early convolutional models used 2D CNNs to learn spatial representations from facial videos \cite{SynRhythm,9607515}, but these approaches had limited ability to capture temporal physiological dynamics. Subsequent methods introduced sophisticated CNN architectures, including PhysNet and DeepPhys, which jointly model spatial and temporal information to better recover pulse signals from short video sequences \cite{physnet,deepphys,electronics13071334}.

Attention mechanisms were later incorporated to improve robustness to motion and irrelevant background regions. Convolutional attention networks and temporal shift attention methods selectively emphasize informative spatial and temporal features associated with physiological activity \cite{TSCAn,stAttention}. More recently, transformer-based architectures have been explored for capturing longer-range temporal dependencies in rPPG signals, including TransRPPG, EfficientPhys, and PhysFormer \cite{transRPPG,efficientphys,physformer}. RhythmMamba further extends this direction using selective state-space modeling for efficient long-sequence physiological representation learning \cite{RhythmMambaFR,mamba}.

Despite these advances, most learning-based rPPG methods have been developed and evaluated on adult datasets captured under relatively constrained conditions. Their applicability to neonatal monitoring remains unclear because neonatal videos exhibit substantially different characteristics, including rapid motion, frequent occlusions, varying head orientations, lower signal-to-noise ratios, and substantially higher heart rates than adults.

\subsection{\texorpdfstring{Camera-based SpO\textsubscript{2} estimation}{Camera-based SpO2 estimation}}

Compared with HR estimation, camera-based estimation of oxygen saturation (SpO\textsubscript{2}) remains substantially under-explored. Estimating SpO\textsubscript{2} from RGB video is more challenging because oxygen saturation depends on subtle wavelength-dependent absorption characteristics and requires reliable estimation of both pulsatile and baseline signal components.

One recent approach uses spatiotemporal representations with an EfficientNet-B3-style regression backbone to estimate SpO\textsubscript{2} from facial videos \cite{cheng2024contactless}. Other methods first derive rPPG representations and subsequently apply temporal models, such as CNN-BiLSTM architectures, for oxygen saturation prediction \cite{clSpO2}. However, these studies primarily focus on adult subjects and are typically evaluated under relatively constrained conditions.

\subsection{Neonatal datasets and contact-free monitoring studies}

Most publicly available rPPG datasets and benchmarks are based on adult subjects, leaving neonatal monitoring comparatively underexplored. A key challenge in neonatal rPPG research is the variability encountered in real clinical environments. Differences in illumination, infant pose, facial visibility, skin tone distribution, and bedside acquisition conditions can introduce substantial domain shifts that affect model generalization. Consequently, evaluating models on a single neonatal dataset may provide an incomplete assessment of robustness and external validity.

Prior neonatal camera-based monitoring studies have primarily focused on feasibility demonstrations and signal-processing-based physiological estimation rather than end-to-end learning approaches, which have shown higher accuracy for adult rPPG. Early NICU studies demonstrated the possibility of continuous non-contact monitoring of HR, respiration, and oxygenation-related signals from video recordings of preterm infants \cite{Villarroel2014ContinuousNonContactVitalSign,Villarroel2019NonContactPhysiologicalMonitoring}. These studies also emphasized persistent practical challenges in neonatal settings, including motion artifacts, occlusion, changing illumination, and inconsistent face visibility. NBHRnet \cite{neonates} is one of the few deep learning approaches specifically designed for neonatal HR estimation and is evaluated on the NBHR dataset. The method combines kernelized correlation filter-based tracking with augmentation strategies to improve robustness. While NBHR represents an important step toward neonatal contact-free monitoring, it reflects a limited acquisition setting and population distribution, and does not address SpO\textsubscript{2} estimation.

More recent work has explored optical and hardware-assisted approaches for oxygenation estimation in pediatric and neonatal populations \cite{Ye:24}. However, these systems depend on calibrated optical setups, handcrafted physiological features, or classical regression techniques rather than learning-based models trained directly from RGB video. As a result, neonatal SpO\textsubscript{2} estimation from standard RGB facial videos remains insufficiently studied.

%% file: methodology.tex
\section{Methodology}

\begin{figure}[t]
    \centering
    \includegraphics[width=\columnwidth]{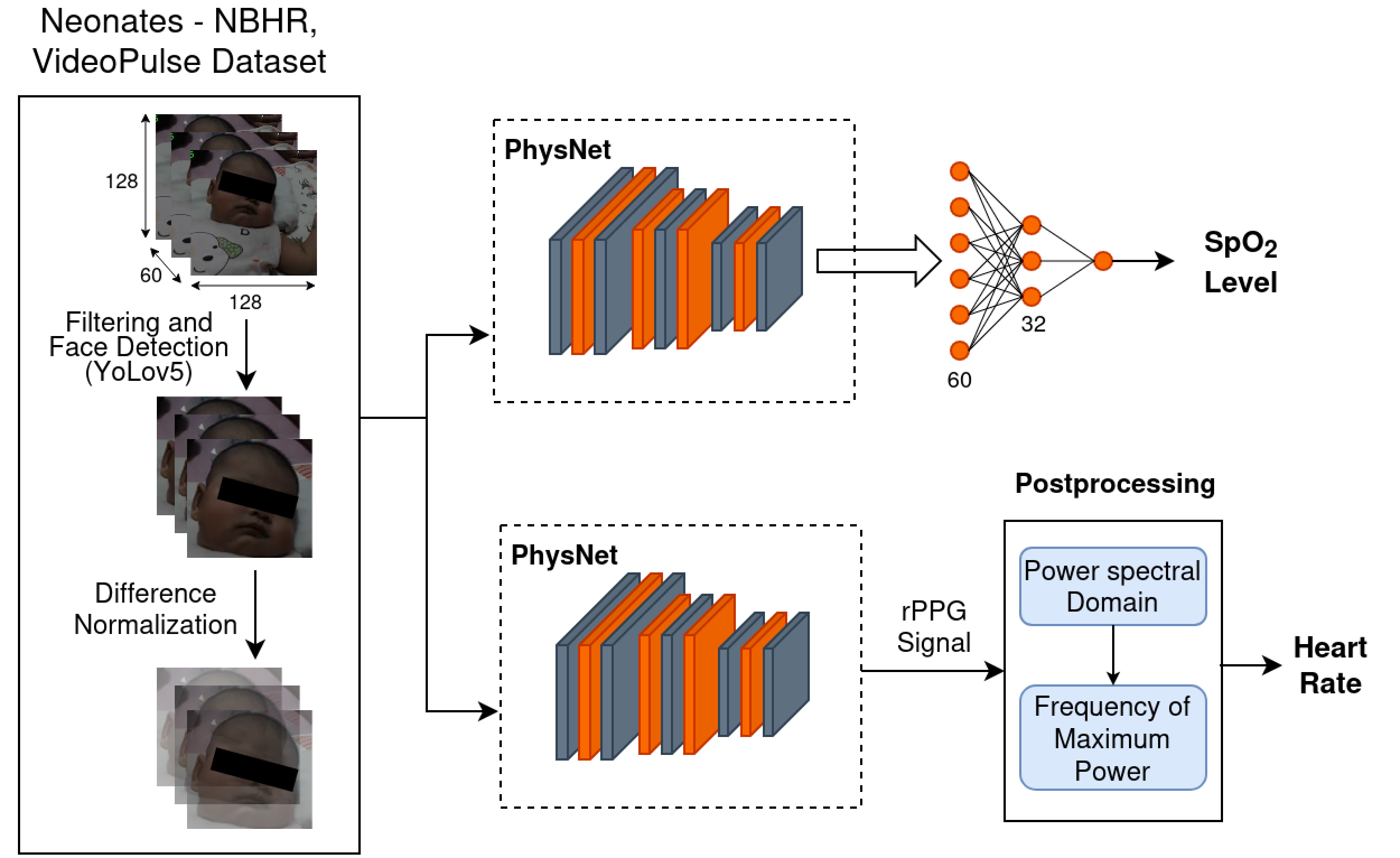}
    \caption{\textbf{System architecture}. The proposed HR and \texorpdfstring{SpO\textsubscript{2}}{SpO2} prediction pipeline is shown. 
    The HR model trained on NBHR was evaluated directly on VideoPulse, whereas the \texorpdfstring{SpO\textsubscript{2}}{SpO2} model pre-trained on NBHR was fine-tuned on VideoPulse. The same model architecture was also trained on the PURE dataset for adult \texorpdfstring{SpO\textsubscript{2}}{SpO2} estimation.}
    \label{fig:methodology_diagram}
\end{figure}

The proposed neonatal HR and \texorpdfstring{SpO\textsubscript{2}}{SpO2} prediction pipeline is shown in Fig.~\ref{fig:methodology_diagram}. Two previously published datasets (NBHR, PURE) \cite{pure, neonates}, and our own novel dataset (VideoPulse) were used for model training and evaluation. The datasets consist of RGB facial videos paired with ground truth data, including HR, PPG waveforms and \texorpdfstring{SpO\textsubscript{2}}{SpO2} level obtained through contact-based measurement devices.

\subsection{Preprocessing}
A shared preprocessing pipeline was applied for both neonatal HR and \texorpdfstring{SpO\textsubscript{2}}{SpO2} estimation. Each input video was segmented into two second clips (60 frames at 30 fps). Unlike adult subjects, who can maintain a stable and consistent pose during video capture, neonates frequently exhibit spontaneous movements, making it challenging to obtain properly aligned facial videos. To address this, a face alignment step was done prior to feeding video frames into the model. For face detection, RetinaFace\cite{Dosso2022NICUfaceRobustNeonatalFaceDetection,Grooby2023NeonatalFaceLandmarkDetection} with a ResNet 50 backbone\cite{Dosso2022NICUfaceRobustNeonatalFaceDetection,Zhao2024DeeperInsightNeonatalFaceDetection} and Haar cascade \cite{haarcascade} were compared against a YOLOv5 based face detector pretrained on adult face datasets. While RetinaFace and Haar cascade failed to perform reliably on neonatal videos, the YOLOv5 detector was used since it produced more stable detections on neonatal faces without additional fine tuning\cite{Dosso2022NICUfaceRobustNeonatalFaceDetection}. The detector attempts to localize the face in the first frame and if unsuccessful, it advances by 30 frames and retries until a bounding box is found. Once detected, the same bounding box is applied to the remaining frames of the 2 second clip.
Since neonatal faces were not always upright, when no face was detected, the video was rotated in $90^\circ$ increments until detection succeeded. The detected facial region was then cropped and resized to $128 \times 128$ pixels. After alignment and cropping, temporal difference normalization \cite{rppgtoolbox} was applied to emphasize subtle pulse related intensity variations while suppressing static features and lighting bias. Differences between consecutive frames were computed and normalized by their standard deviation, producing a standardized input for the deep learning model.

\subsection{Heart Rate Prediction}
Neonatal HR prediction was performed using the PhysNet 3D CNN trained on the NBHR neonatal dataset. 

\subsubsection{Ground truth PPG denoising}
\label{subsec:filtering}

The acquisition of clean ground truth PPG data is particularly challenging for neonates since they are often quite active during data collection, leading to significant motion-induced artifacts in the PPG signals. Although pulse oximeter probes specially designed for neonates were used in both the NBHR and VideoPulse datasets, there was a high level of noise due to motion in the recorded ground truth signals compared to adult datasets.


To address this issue, a PPG reconstruction pipeline was developed based on established Generative Adversarial Network (GAN) architectures for signal denoising. Low quality segments of the ground truth PPG waveform were identified using a one class Support Vector Machine (SVM) which had been previously used for PPG signal quality assessment~\cite{feli2023end, feli2023energy}. This pretrained SVM model and inference procedure was applied to our recordings using sliding window segmentation with 30 second windows and a 2 second shift, resulting in a 28 second overlap between consecutive windows (about 93.3\% overlap). Noisy segments shorter than a predefined threshold (15 seconds) were then processed using a pretrained GAN based reconstruction model~\cite{wang2022ppg, kazemi2022robust, feli2023end}. The pretrained generator and discriminator were used for reconstruction, where the generator estimates cleaned windows and the window is advanced by the same 2 second shift for iterative reconstruction. This supports denoising while preserving HR related features such as peak timing and waveform shape.

\begin{figure*}[t]
    \centering

    \begin{subfigure}{0.3\textwidth}
        \centering
        \includegraphics[width=\linewidth]{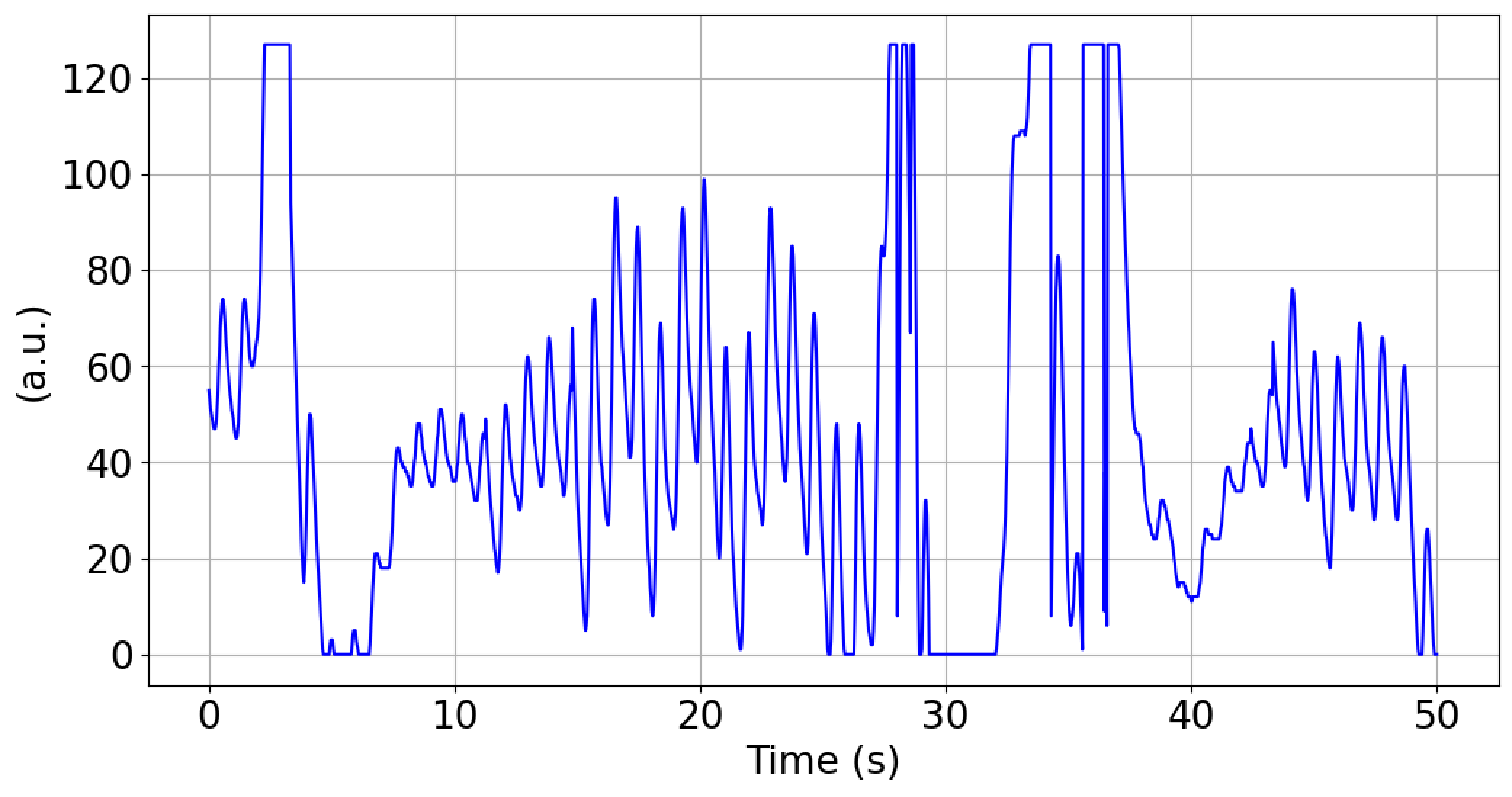}
        \caption{}
    \end{subfigure}
    \hfill
    \begin{subfigure}{0.3\textwidth}
        \centering
        \includegraphics[width=\linewidth]{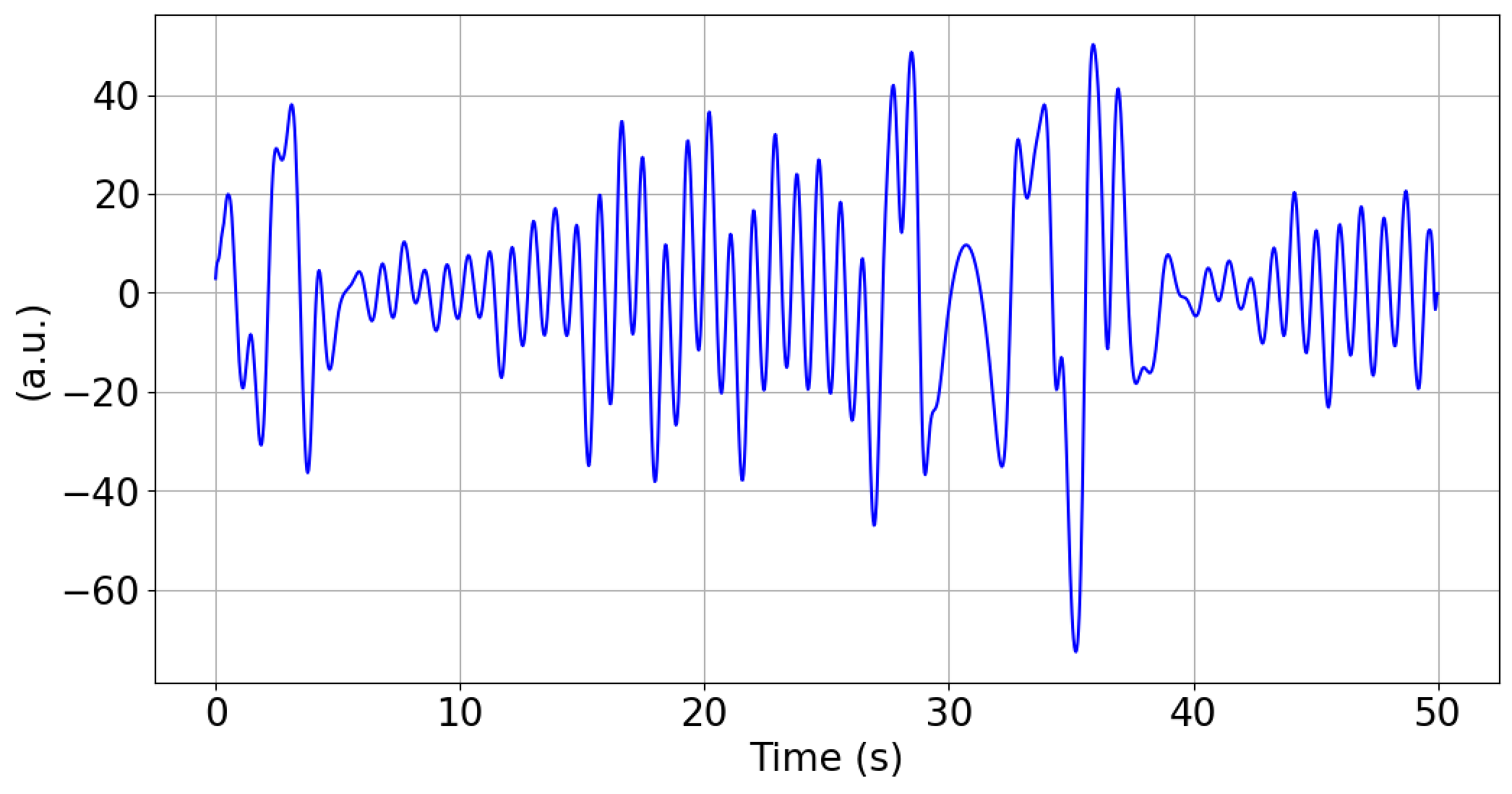}
        \caption{}
    \end{subfigure}
    \hfill
    \begin{subfigure}{0.3\textwidth}
        \centering
        \includegraphics[width=\linewidth]{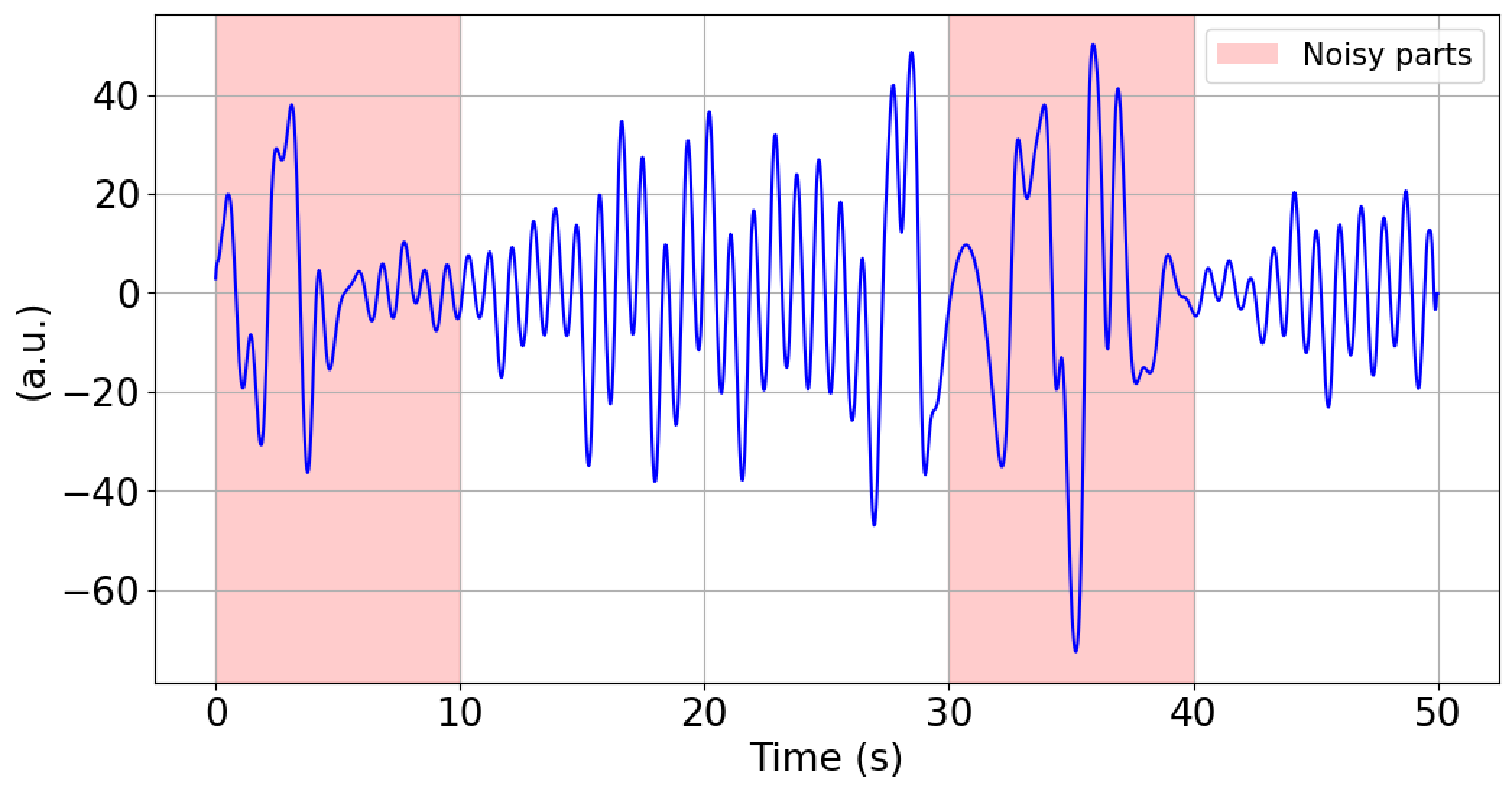}
        \caption{}
    \end{subfigure}

    \caption{\textbf{Ground truth PPG denoising.} \textbf{(a):} Raw PPG signals from the oximeter containing motion artifacts. These are first classified by an SVM to identify noisy regions. \textbf{(b):} The reconstructed signal after the noisy segments were reconstructed using a GAN. \textbf{(c):} The reconstructed signal is further filtered to recover clean PPG segments for reliable rPPG analysis. The red shaded regions are removed since they are too noisy.}

    \label{fig:reconstruction_diagram}
\end{figure*}

Following reconstruction a filtering algorithm was applied to further refine the PPG signal, as shown in Fig.~\ref{fig:reconstruction_diagram}. Heart rate variability (HRV) was analyzed within sliding windows of 2 seconds. If a window exhibited abnormal variability, defined as a fluctuation greater than 15 bpm, it was marked as noisy and excluded from the final training and evaluation set. The corresponding 2-second video clips were also discarded to ensure consistency between videos and ground-truth signals. 



\subsubsection{Loss function}
The negative Pearson correlation loss between the predicted rPPG waveform and the ground truth PPG waveform was used during training. For each input clip, let $s_{\text{pre}}, s_{\text{gt}}\in\mathbb{R}^{T}$ denote the predicted and ground truth waveforms over $T$ temporal samples. The loss is given by
\begin{equation}
L_p = 1 - \frac{\sum_{t=1}^{T} (s_{\text{pre},t}-\bar{s}_{\text{pre}})(s_{\text{gt},t}-\bar{s}_{\text{gt}})}
{\sqrt{\sum_{t=1}^{T}(s_{\text{pre},t}-\bar{s}_{\text{pre}})^2}\sqrt{\sum_{t=1}^{T}(s_{\text{gt},t}-\bar{s}_{\text{gt}})^2}+\epsilon},
\label{eq:pearson_loss}
\end{equation}
where $\bar{s}_{\text{pre}}$ and $\bar{s}_{\text{gt}}$ are temporal means and $\epsilon$ is a small constant for numerical stability, and is averaged over clips in the minibatch.





\subsubsection{Post-processing}

Postprocessing steps were then applied to extract the HR from the predicted rPPG signal. According to prior research, there are two primary methods \cite{9607515, physformer} for extracting HR from PPG or rPPG signals: peak detection or determining dominant frequencies using the power spectral density (PSD).  
The latter PSD method was employed for HR extraction. Initially, a second-order Butterworth bandpass filter with a cutoff range of 0.4 Hz to 4 Hz was applied to the rPPG signal to isolate the relevant frequency components. Subsequently, a Fast Fourier Transform (FFT) was performed on the filtered signal to compute its power spectral density. The HR was then extracted as the frequency component with the highest power.

\subsection{Blood Oxygen Saturation Prediction}



\subsubsection{Modified PhysNet Model for \texorpdfstring{SpO\textsubscript{2}}{SpO2} Prediction}

The PhysNet model was adapted for \texorpdfstring{SpO\textsubscript{2}}{SpO2} regression by appending a three layer fully connected head to the final PhysNet representation. Specifically, two hidden layers of 60 and 32 neurons are used with ReLU activations, followed by a single neuron linear output layer for continuous \texorpdfstring{SpO\textsubscript{2}}{SpO2} prediction.  The PhysNet model was trained using the NBHR dataset after incorporating the filtering step mentioned in Subsection~\ref{subsec:filtering}. The same preprocessing pipeline used for HR prediction was applied to the \texorpdfstring{SpO\textsubscript{2}}{SpO2} prediction task to ensure consistency in data preparation. For VideoPulse SpO\texorpdfstring{$_2$}{2} estimation, the model was initialized using a pre-trained checkpoint obtained from training on the NBHR dataset. To retain the general spatiotemporal representations learned from the NBHR dataset, the first two 3D convolutional layers of the model were frozen during fine-tuning. The remaining layers were then fine-tuned using the VideoPulse training partition.
\subsubsection{Optimizations for neonatal \texorpdfstring{SpO\textsubscript{2}}{SpO2} prediction}
The NBHR dataset exhibits a right-skewed distribution (more \texorpdfstring{SpO\textsubscript{2}}{SpO2} values near 100\%) with notable class imbalance. Imbalanced target distributions are common for \texorpdfstring{SpO\textsubscript{2}}{SpO2} in clinical data where labels cluster near high saturation values. 
To mitigate the effects of class imbalance, Label Distribution Smoothing (LDS)~\cite{lds} was applied using a Beta kernel with kernel size $k_s = 7$ and shape parameters $\alpha = 2$ and $\beta = 5$, resulting in a smoother and more continuous approximation of the empirical label distribution. The resulting smoothed distribution was used to compute sample-wise weights that guide the learning process. To further enhance generalization, each video input sequence was reversed along the temporal dimension and then fed into the model as an additional training sample. This augmentation enables the model to capture temporal patterns more robustly by learning from both the original and time-reversed sequences.



\subsubsection{Loss function}
Normal saturation values occur far more often than low saturation values in the training data, and hence optimizing standard RMSE can under-emphasize errors in rare but clinically relevant ranges. Hence a weighted RMSE objective was used during training to mitigate class imbalance. Sample weights were computed from the smoothed label distribution (LDS) estimated on the training set, assigning larger weights to rarer labels. The loss is given by
\begin{equation}
L_{\text{wRMSE}} = \sqrt{\frac{\sum_{i=1}^{N} w_i \left( s_{\text{pre},i} - s_{\text{gt},i} \right)^2}{\sum_{i=1}^{N} w_i + \epsilon}},
\label{eq:weighted_rmse}
\end{equation}
where $N$ is the number of samples in a mini batch, $s_{\text{gt},i}$ and $s_{\text{pre},i}$ are the ground truth and predicted \texorpdfstring{SpO\textsubscript{2}}{SpO2} values for sample $i$, and $w_i$ is the LDS derived weight for the corresponding label. The weights are normalized so that their mean is 1, which preserves the scale of the loss. 

\subsubsection{\texorpdfstring{SpO\textsubscript{2}}{SpO2} Prediction on an Adult Dataset}

Since our pipeline is one of the first to investigate \texorpdfstring{SpO\textsubscript{2}}{SpO2} prediction from facial video, we further evaluated it on the well-known PURE dataset~\cite{pure} to assess its performance on adult rPPG \texorpdfstring{SpO\textsubscript{2}}{SpO2} prediction as well. For this experiment, face detection was performed using the Haar Cascade method. The PURE dataset was divided by participant into mutually exclusive training, validation, and test partitions using a 70\%/10\%/20\% ratio. The same model architecture used in the neonate experiments was employed for this evaluation. 


\subsection{Implementation details}


The NBHR dataset was divided by neonate into mutually exclusive training, validation, and test partitions using a 70\%/15\%/15\% ratio. For NBHR, VideoPulse, and PURE, the participant was the partitioning unit: all recordings from the same participant were retained exclusively in one partition. For HR, the NBHR-trained model was evaluated directly on the VideoPulse dataset without VideoPulse fine-tuning. For \texorpdfstring{SpO\textsubscript{2}}{SpO2}, the NBHR-pretrained model was fine-tuned using the VideoPulse training partition and evaluated only on the held-out VideoPulse test partition. Training was performed on a system equipped with an NVIDIA RTX 4090 GPU. The models were trained for 27 epochs using an initial learning rate of 0.01. To improve convergence and stabilize training, the OneCycleLR learning rate scheduler was employed, allowing the learning rate to vary dynamically during training. All preprocessing steps, including face detection, alignment, difference normalization, and filtering, were applied consistently across training, validation, and testing datasets to ensure fair evaluation and reproducibility. 




%% file: datasets.tex
\section{Datasets}
\label{dataset}
\subsection{Dataset Overview}
We used the NBHR and our own VideoPulse rPPG neonatal datasets for training and testing. Our model pipeline was also trained and tested on the PURE~\cite{pure} adult rPPG dataset. Brief descriptions of the datasets used in this study are given below.


\emph{NBHR} consists of 1130 videos and reference vital signs that are recorded from 257 infants at 0--6 days old. The dataset includes facial videos and corresponding synchronized physiological signals including PPG information, HR and SpO\textsubscript{2} level. 

\emph{PURE} is a benchmark rPPG dataset consisting of 10 adult subjects recorded over 6 sessions. Each session lasted approximately 1 minute, and raw video was recorded at 30 fps. The 6 sessions for each subject consisted of: (1) steady, (2) talking, (3) slow head translation, (4) fast head translation, (5) small, and (6) medium head rotations. Pulse rates are at or close to the subject's resting rate. In addition to video, the dataset provides synchronized PPG waveforms, SpO\textsubscript{2}, HR ground truths.

\subsection{The VideoPulse dataset}

\emph{VideoPulse} consists of 157 video recordings collected from 52 neonates (postnatal age 0 to 6 days) in the postnatal ward. The cohort included 25 male and 27 female neonates, with mean postnatal age 4.77 days.
The VideoPulse data-collection protocol was approved by the
University Ethics Review Committee, University of Moratuwa,
and the Institutional Ethical Review Committee of De Soysa
Maternity Hospital for Women. 
Written informed consent was obtained from a
parent or legal guardian of each neonatal participant. 

Data were collected using an overhead, downward facing Logitech C920 HD Pro webcam (1080p, 30 fps, 78 degree diagonal field of view, autofocus) mounted on a tripod with a horizontal boom arm, together with a neonatal pulse oximeter. To reduce reliance on upright face alignment and to better reflect real-world conditions, recordings were taken across three predefined orientations. The total video duration was 2.62 hours, with the average video duration being 60 seconds. There were 2 or 3 videos of each recording scenario. Both devices were connected to a single PC to acquire time synchronized facial video and pulse oximeter PPG signals. Reference PPG waveforms were recorded using a Contec CMS60D pulse oximeter. The mean HR was 113.99 bpm (std. dev. = 13.96 bpm, maximum = 175 bpm, minimum = 79 bpm). The mean SpO\textsubscript{2} was 94.45\% (std. dev. = 2.80 \%, maximum = 99 \%, minimum = 87 \%). A custom \texttt{C++} application was used to simultaneously record the raw video feed from the camera and the ground truth signals, ensuring that both were temporally aligned. The videos were recorded at 30 fps, while the ground truth data was sampled at 60 Hz. 




%% file: results.tex
\section{Results}

The following metrics were used for evaluation: mean absolute error (MAE), root mean square error (RMSE), mean absolute percentage error (MAPE), and standard deviation of the error (SD). The prediction time window length (TW) is also reported to indicate latency. For HR, errors are expressed in beats per minute (bpm), and for \texorpdfstring{SpO\textsubscript{2}}{SpO2}, errors are expressed in percent saturation (\%).

\subsection{Neonatal Heart Rate Estimation}

The proposed approach is evaluated against several state-of-the-art methods in neonatal HR prediction from videos, including three data-driven approaches and three signal-processing--based methods, namely: CHROM, 2SR, POS, EVM-CNN, PRNet and NBHRnet. Our method demonstrated the best performance compared to both data-driven deep learning models and traditional signal processing approaches in the task of neonatal HR prediction as shown in Table~\ref{tab:hr_all} when testing on the NBHR dataset. Notably, our model achieved the best accuracy while using shorter time window lengths compared to the state-of-the-art. For the NBHR dataset our method achieved a best performance MAE of 2.80 bpm and MAPE of 2.44\% for the 6 s window and the lowest value for SD of 5.12 bpm for the 4 s window. This suggests that our approach not only improves prediction accuracy but also reduces latency in HR monitoring, which is especially valuable in neonatal care settings where timely and accurate monitoring is critical.

\begin{table}[t]
  \centering
  \caption{Heart rate estimation comparison}
  \label{tab:hr_all}
  \footnotesize
  \setlength{\tabcolsep}{12pt}
  \renewcommand{\arraystretch}{1.25}
  \begin{tabular}{@{}lcccc@{}}
    \toprule
    \textbf{Method} & \textbf{TW} & \textbf{MAE} & \textbf{SD} & \textbf{MAPE} \\
                    & \textbf{(s)}$\downarrow$ & \textbf{(bpm)}$\downarrow$ & \textbf{(bpm)}$\downarrow$ & \textbf{(\%)}$\downarrow$ \\
    \midrule
    \textit{NBHR dataset} & & & &\\ 
CHROM~\cite{dehaan2013chrom} & 12 & 6.34 & 8.02 & 5.49 \\
2SR~\cite{wang2016_2sr} & 6 & 5.48 & 7.69 & 4.93 \\
POS~\cite{7565547} & 10 & 5.96 & 10.93 & 5.12 \\
EVM-CNN~\cite{qiu2019_evmcnn} & 8 & 4.51 & 5.78 & 3.91 \\
PRNet~\cite{huang2021_prnet} & 2 & 5.66 & 7.83 & 4.74 \\
NBHRnet~\cite{neonates} & 2 & 3.97 & 5.32 & 3.28 \\
NBHRnet~\cite{neonates} & 4 & 3.83 & 5.16 & 3.20 \\
NBHRnet~\cite{neonates} & 6 & 3.76 & 5.15 & 3.13 \\
    \addlinespace
    Ours  & 2 & 2.97 & 5.36 & 2.64 \\
    Ours  & 4 & 2.89 & \textbf{5.12} & 2.45\\
    Ours  & 6 & \textbf{2.80} & 5.23 & \textbf{2.44} \\
    Ours  & 8 & 3.23 & 5.19 & 2.83 \\ \midrule
    \textit{VideoPulse dataset} & & & &\\ 
    Ours & \textbf{2} & \textbf{5.34} & \textbf{7.12} & \textbf{4.32}\\
    \bottomrule
  \end{tabular}
\end{table}

\begin{figure}[h]
    \begin{subfigure}[b]{0.48\columnwidth}
        \centering
        \includegraphics[width=\linewidth]{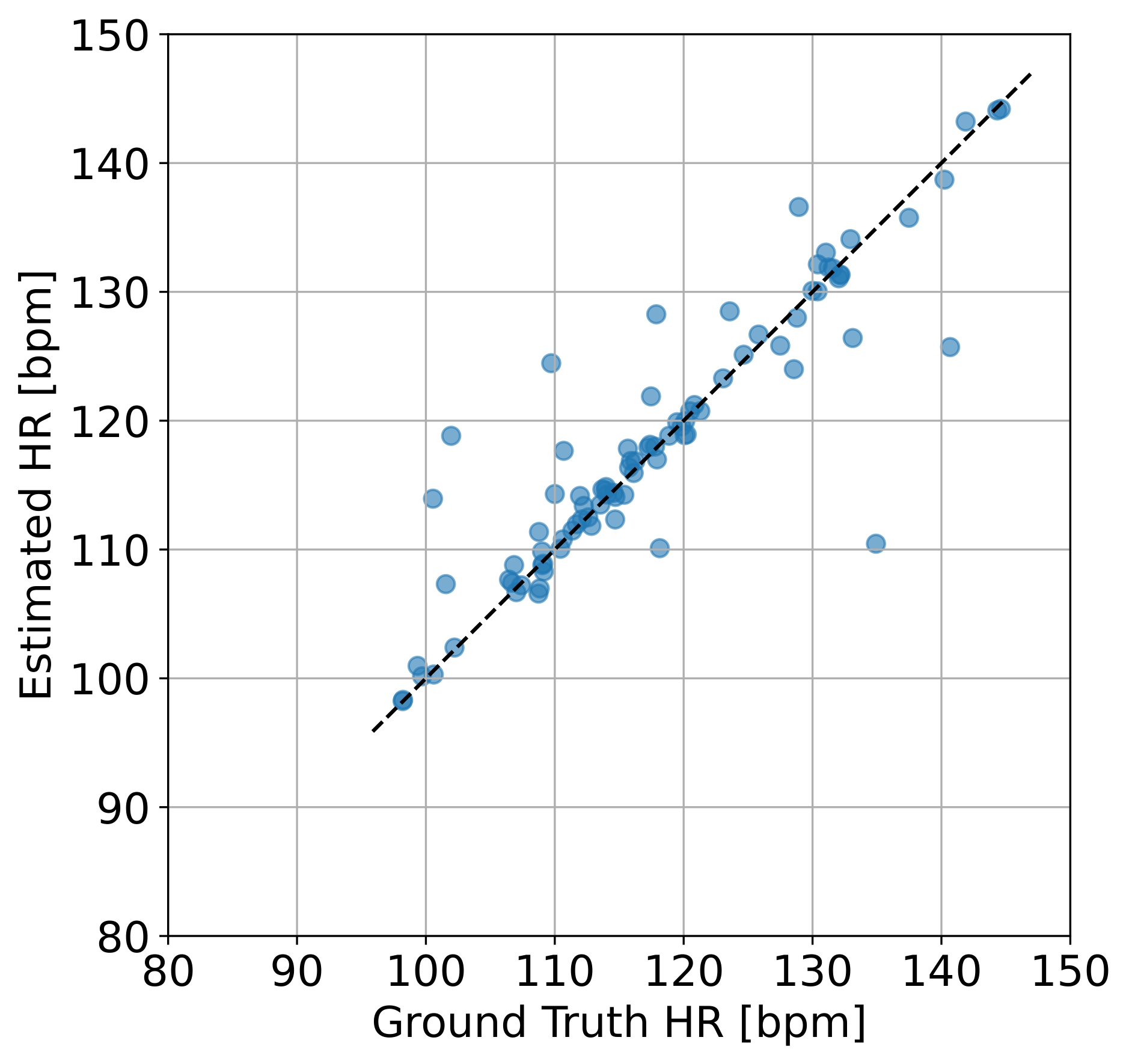}
        \caption{}
    \end{subfigure}
    \begin{subfigure}[b]{0.48\columnwidth}
        \centering
        \includegraphics[width=\linewidth]{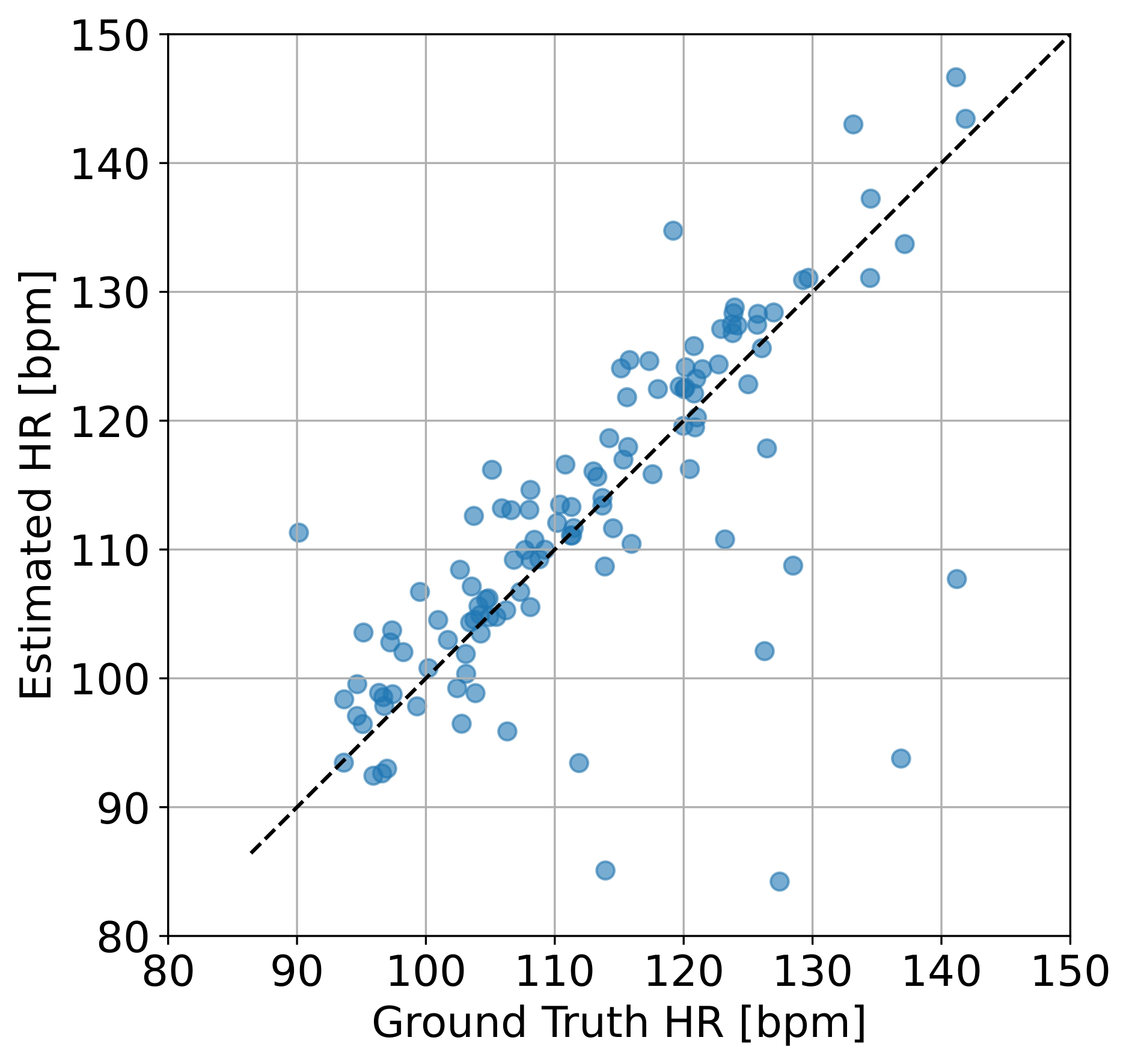}
        \caption{}
    \end{subfigure}

    \caption{\textbf{Neonatal heart rate estimation results. (a): }Evaluation on the test split of NBHR dataset. \textbf{(b): }Evaluation on the test split of VideoPulse dataset. Predictions align closely with the identity line.}
    \label{fig:nbhr_evaluation}
\end{figure}

In addition, our method obtained reasonable results on the VideoPulse dataset. For HR, the NBHR-trained model was evaluated on the VideoPulse dataset without VideoPulse fine-tuning. For \texorpdfstring{SpO\textsubscript{2}}{SpO2}, the NBHR-pretrained model was fine-tuned using the VideoPulse training partition and evaluated on the held-out VideoPulse test partition. The VideoPulse partitions were participant-disjoint, so no recordings from a test participant appeared in the fine-tuning data. Table~\ref{tab:hr_all} summarizes HR perfomance, while Table~\ref{tab:nbhr} reports \texorpdfstring{SpO\textsubscript{2}}{SpO2} performance on the VideoPulse and NBHR datasets.
As shown in Fig.~\ref{fig:nbhr_evaluation}, the HR predictions of the proposed model closely track the ground truth, achieving Pearson correlation coefficients of 0.92 and 0.88 on the NBHR and VideoPulse test sets, respectively.

\subsection{Neonatal \texorpdfstring{SpO\textsubscript{2}}{SpO2} Estimation}


\subsubsection{Neonatal \texorpdfstring{SpO\textsubscript{2}}{SpO2} prediction}

\begin{figure}
    \begin{subfigure}[b]{0.48\columnwidth}
        \centering
         \includegraphics[width=\linewidth]{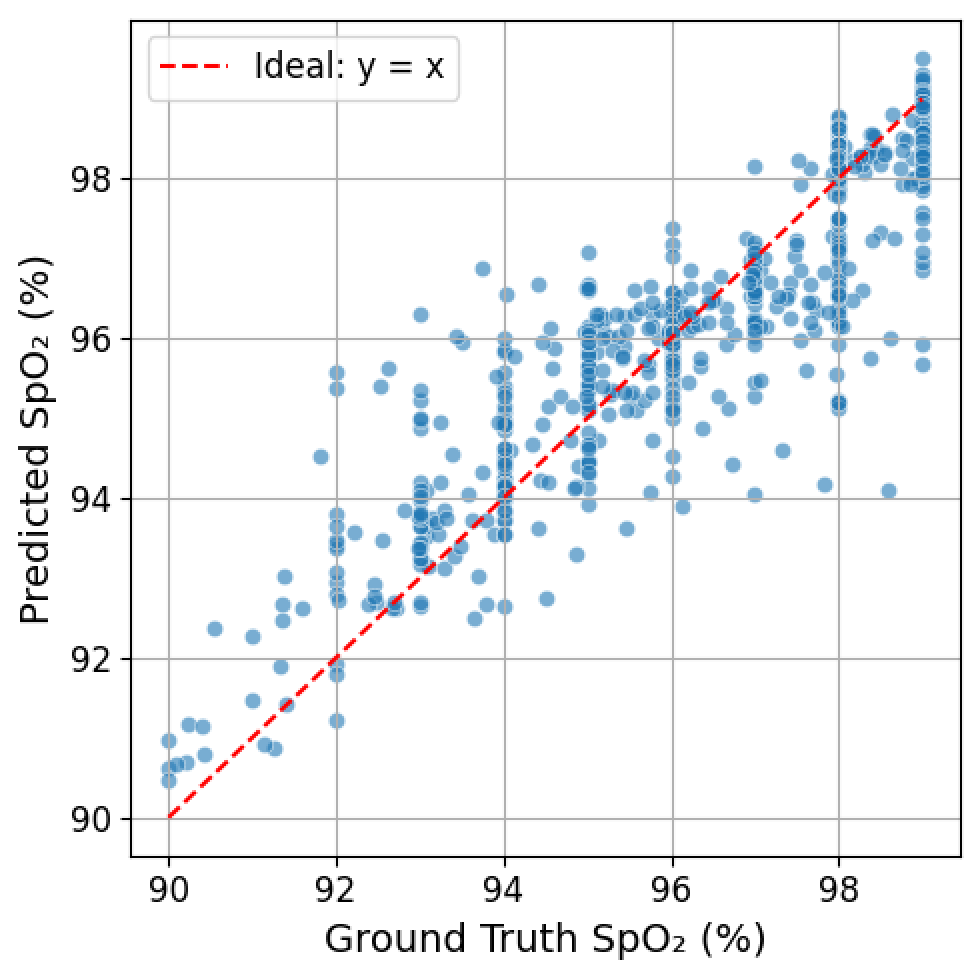}
        \caption{}
    \end{subfigure}
    \begin{subfigure}[b]{0.48\columnwidth}
        \centering
        \includegraphics[width=\linewidth]{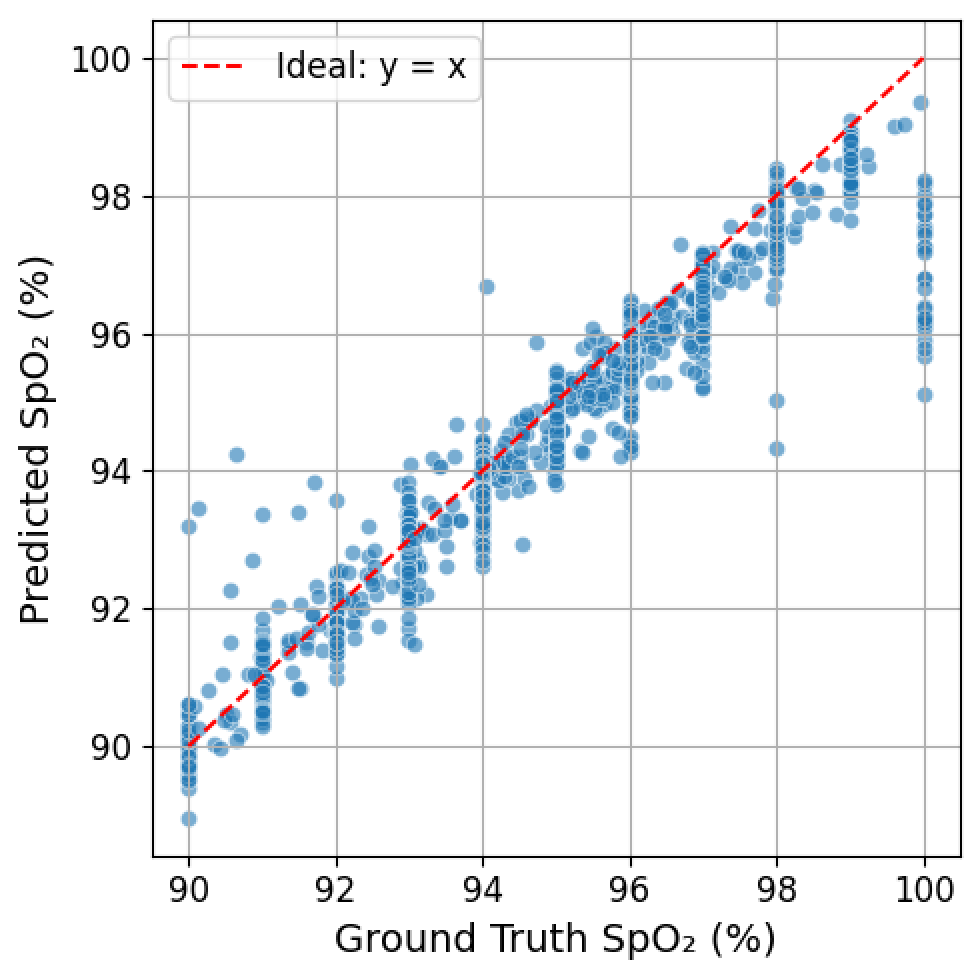}
        \caption{}
    \end{subfigure}

    \caption{\textbf{Neonatal \texorpdfstring{SpO\textsubscript{2}}{SpO2} estimation results. (a): }Evaluation on the test split of NBHR dataset. \textbf{(b): }Evaluation on the test split of \textit{VideoPulse} dataset. Predictions align closely with the identity line in most cases, though there are some outliers near 100\% saturation for \textit{VideoPulse} dataset.}
    \label{fig:scatter_spo2}
\end{figure}

To the best of our knowledge, this work constitutes the first attempt to predict \texorpdfstring{SpO\textsubscript{2}}{SpO2} levels in neonates using a video-based deep learning approach. First, we evaluate our modified PhysNet model on the publicly available NBHR dataset. Subsequently, we fine-tune the model using the participant-disjoint VideoPulse training partition and evaluate it on the held-out VideoPulse test partition. As shown in Fig.~\ref{fig:scatter_spo2}, the predicted \texorpdfstring{SpO\textsubscript{2}}{SpO2} closely follows the reference values on both NBHR and VideoPulse datasets, obtaining Pearson correlation coefficients of 0.85 and 0.93 on NBHR and VideoPulse, respectively. Table~\ref{tab:nbhr} presents the performance of our method on both datasets. When compared with prior work on neonates reported in \cite{Ye:24}, where a clinical trial was performed on 22 neonates using a specialized notch RGB camera under controlled conditions, our approach has improved MAE. 

\begin{table}[t]
  \centering
  \caption{Neonatal \texorpdfstring{\texorpdfstring{SpO\textsubscript{2}}{SpO2}}{SpO2} estimation results.}
  \label{tab:nbhr}
  \setlength{\tabcolsep}{5pt}
  \small
  \begin{tabularx}{\columnwidth}{@{}Y Y cc@{}}
    \toprule
    \textbf{Method} &
    \textbf{Dataset} &
    \textbf{MAE(\%)$\downarrow$ } &
    \textbf{RMSE(\%)$\downarrow$} \\
    \midrule
    Ye et. al. \cite{Ye:24} &
    (private dataset) &
    3.41 to 3.17 &
    - \\
    Ours & NBHR & 1.69 & 2.20 \\
    Ours & VideoPulse & \textbf{1.68} & \textbf{2.18} \\
    \bottomrule
  \end{tabularx}
\end{table}

\subsubsection{Ablation Study}

An ablation study (see Table~\ref{tab:ablation-nbhr}) was conducted on the NBHR dataset to assess the effect of different training strategies. The architecture (PhysNet 3D-CNN), optimizer, training budget, and NBHR data splits were held constant and only the training strategy was varied. Using only RMSE as the loss for training yields an \texorpdfstring{SpO\textsubscript{2}}{SpO2} RMSE of 2.74\%, and this is considered the baseline. Replacing it with the label-distribution--smoothed (LDS) weighted RMSE reduces the RMSE to 2.30\% (-0.44 absolute change; -16.1\% change vs. baseline). Adding time-reversal augmentation further reduced RMSE to 2.20\% (additional -0.10; -19.7\% vs. baseline). Since NBHR labels are concentrated around normal saturation values, training using only RMSE tends to prioritize the majority region and can bias errors on less frequent values. LDS weighted RMSE rebalances the contribution of each label region using a smoothed estimate of label frequency, which reduces this imbalance driven bias and improves overall RMSE. Time reversal further acts as a regularizer, because the rPPG waveform is approximately quasi periodic over short windows, and reversing the sequence discourages reliance on direction specific temporal artifacts such as motion trends, thereby improving generalization.

\begin{table}[t]
  \centering
  \caption{Ablation study on NBHR for \texorpdfstring{SpO\textsubscript{2}}{SpO2} model}
  \label{tab:ablation-nbhr}
  \small
  \setlength{\tabcolsep}{5pt}
  \renewcommand{\arraystretch}{1.05}
  \begin{tabularx}{\columnwidth}{@{}Y
                                 S[table-format=1.2]
                                 S[table-format=1.2]@{}}
    \toprule
    \textbf{Training strategy} & \textbf{RMSE(\%)$\downarrow$} & \textbf{MAE(\%)$\downarrow$} \\
    \midrule
    Standard RMSE loss & 2.74 & 1.96 \\
    Weighted RMSE (LDS) & 2.30 & 1.77 \\
    Weighted RMSE plus time reversal augmentation & \textbf{2.20} & \textbf{1.69} \\
    \bottomrule
  \end{tabularx}
\end{table}

\subsection{Adult \texorpdfstring{SpO\textsubscript{2}}{SpO2} Estimation}

Table~\ref{tab:pure} shows the performance of different models on the PURE dataset for adult \texorpdfstring{SpO\textsubscript{2}}{SpO2} estimation. The citations identify the originating methods or architectures; the listed RMSE values were obtained in our experiments and are
not values quoted from those papers. The modified PhysNet architecture with 3D convolutional layers outperformed all other state-of-the-art models evaluated on the PURE dataset in this study, achieving an RMSE of 0.96. Although the focus of this work is neonatal prediction, this result is reported to highlight the suitability of our approach even for adult \texorpdfstring{SpO\textsubscript{2}}{SpO2} estimation.

\begin{table}[t]
  \centering
  \caption{Author-implemented \texorpdfstring{SpO\textsubscript{2}}{SpO2} models evaluated on PURE dataset}
  \label{tab:pure}
  \small
  \setlength{\tabcolsep}{5pt}
  \renewcommand{\arraystretch}{1.05}
  \begin{tabularx}{\columnwidth}{@{}Y l S[table-format=2.2]@{}}
    \toprule
    \textbf{Model} &
    \textbf{Backbone} &
    {\textbf{RMSE(\%)}$\downarrow$} \\
    \midrule
    STMap + EfficientNet-B3~\cite{cheng2024contactless}
      & 2D CNN & 26.80 \\
    RhythmMamba adapted for \texorpdfstring{SpO\textsubscript{2}}{SpO2}~\cite{RhythmMambaFR}
      & Mamba & 18.43 \\
    Modified PhysNet (ours)
      & 3D CNN & \bfseries 0.96 \\
    \bottomrule
  \end{tabularx}
\end{table}


%% file: discussion.tex
\section{Discussion}

This work demonstrates that neonatal heart rate and SpO$_2$ can be estimated from short facial video segments using a deep learning based rPPG pipeline. Accurate predictions from 2-second windows suggest that meaningful physiological information can be extracted even under motion, pose variation, and illumination changes commonly observed in neonatal settings. The improvements in SpO$_2$ estimation further highlight the importance of addressing label imbalance and noisy supervision, where label distribution smoothing and weighted regression improve generalization. Additionally, consistent performance across the NBHR and VideoPulse datasets indicates that the model learns transferable physiological representations, supporting the feasibility of contact-free monitoring in realistic clinical environments.

Despite these promising results, several limitations remain. The dataset size is still relatively small, recordings were collected using a single acquisition setup, and pulse oximeter ground-truth signals may contain motion artifacts. These factors may limit generalization across broader clinical settings. Future work should focus on expanding multi-center neonatal datasets, improving robustness to motion and illumination, and evaluating continuous real-time deployment. Incorporating improved face tracking, multi-region analysis, and multimodal sensing may further enhance reliability and move contact-free neonatal monitoring closer to practical clinical use.

%% file: conclusion.tex
\section{Conclusion}
This work introduces VideoPulse, the first end-to-end pipeline that jointly estimates neonatal HR and SpO\textsubscript{2} from video. To our knowledge, it is also the first work that used video-based deep learning for SpO\textsubscript{2} prediction in neonates.
By combining YOLO-guided face alignment, GAN-based signal reconstruction, and label-distribution smoothing with a weighted-RMSE loss, the system demonstrates the feasibility of reliable non-contact monitoring for neonates. The model surpasses state-of-the-art models on public datasets, and also succeeds on our own VideoPulse dataset under varying conditions.
Importantly, the model produces accurate predictions from very short time windows,  
thus highlighting its potential for real-time use in resource-constrained neonatal intensive-care units. By moving beyond traditional HR-only approaches and demonstrating that neonatal SpO\textsubscript{2} can also be estimated using contact-free video, VideoPulse advances the field toward the practical deployment of camera-based neonatal vital-sign monitoring in clinical settings.

%% file: acknowledgement.tex
\section*{Acknowledgment}

We thank Dr. Sheshan Mendis, Dr. M. H. M. Amjad, and the postnatal ward staff at De Soysa Maternity Hospital for their assistance with participant handling and clinical safety supervision during data acquisition.